# How international research teams respond to disruption in their mobility patterns

Olivier J. Walther[1]*, Rafael Prieto-Curiel[2], Erica Odera[1]

[1]University of Florida, [2]Complexity Science Hub Vienna

*Corresponding author: Dr. Olivier J. Walther, Department of Geography, University of Florida, 3131 Turlington Hall, Gainesville, FL 32611, United States, owalther@ufl.edu

11/14/2023

## Abstract

Combining social network analysis with personal interviews, the paper examines how the social structure and internal composition of three Africa-focused international research networks contributes to their resilience. It shows that research networks are structured around a small number of highly influential coordinators. This structure facilitates information exchange and trust between countries and across fields. The study also suggests that the surveyed teams tend to exchange information or trust each other irrespective of their social and professional attributes, indicating that diversity is key to understanding their responses to major shocks such as the COVID-19 pandemic. In a second part, the paper analyzes how the spatial constraints imposed by distance and borders affect their ability to function internationally. It shows that the probability of exchanging information, trusting each other, and co-publishing decreases considerably with distance and that research communities are more likely formed inside the same country than internationally. Interviews reveal that teams responded to travel bans and border closure by emphasizing what they already did best, suggesting that resilience should be considered as an evolutionary attribute of a system.

## Keywords

International research networks; resilience; mobility; COVID-19: Africa; United States

## Acknowledgements

Funding for this research is provided by the National Science Foundation (NSF) EAGER grant #NNFQH1JAPEP3, "Inclusiveness and Diversity as Building Blocks of Resilient International Research Teams in the Age of COVID-19". The authors would like to thank Susanna Goewey, Greg Kiker, Igor Linkov, July Nelson, Jason Padron, Stefani Wald, and the members of the GALUP, SRG and LSIL teams interviewed in this paper.

## 1. Introduction

The last two decades have been characterized by a rise in research networks that cross university, disciplinary and international boundaries (Jones et al., 2008, Adams, 2012; Graf and Kalthaus, 2018). This rapid increase in the demand for scientific collaborations is fueled by the growing complexity of knowledge, the specialization of research fields, competition for major funding, and the emergence of global issues, such as climate change or the recent pandemic, that can only be tackled by multidisciplinary and/or international teams (Wuchty et al., 2007). Research teams are now ubiquitous in the natural and social sciences, as evidenced by the proliferation of multinational research consortiums and co-authored publications (Henriksen, 2016). Of particular interest is the question of how such teams manage to develop international collaborations when confronted with major disruptions in their mobility patterns. Addressing this question requires studying the spatiality of social networks (Glückler et al., 2017; Andris and Sarkar, 2022); in other words, *how* and *where* scientists interact to produce new ideas.

In this exploratory paper, we argue that two main factors shape the resilience of international research networks, defined as "a system's ability to adjust its activity to retain its basic functionality when errors, failures and environmental changes occur" (Gao et al. 2016: 307). The first factor is the architecture of the networks itself, which provides both opportunities and constraints to the individuals involved in scientific collaborations (Tindall, 2015). The second factor that can affect the resilience of international research networks is their geography, or more precisely, how scientists can keep developing new ideas without being physically co-present. The goal of this paper is to examine these factors, by mapping how the social structure of international research networks contributes to their resilience, and how the spatial constraints imposed by distance and borders affect their ability to function internationally.

The resilience of international teams is addressed in light of the recent COVID-19 pandemic that significantly disrupted research activities (Buitendijk et al. 2020), especially those that focused on developing countries, where *ad hoc* transitions to online conferencing did not allow basic communications to continue. For this reason, this paper focuses on three Africa-focused research teams that proved able to recover rapidly from the shock of the pandemic and adapt to new working conditions. Instead of assessing changes over time by testing the same research networks before and after the Covid pandemic, we adopt a mixed-method approach that combines social network analysis (SNA) with personal interviews conducted with key scientists. Since our goal is primarily to identify the conditions that make international research networks resilient to external shocks such as the Covid pandemic, we rely on qualitative information to assess temporal changes. Building on similar studies (Walther et al., 2021), we argue that combining a formal approach to networks with a qualitative assessment of their evolution is well adapted to study both the social and spatial structure of international research networks and, eventually, understand which strategies were developed to cope with the pandemic.

The paper is structured as follows. The next section leverages the literature on the geography of international research networks. The third section discusses our data and methodology. The results of our mixed-method approach are presented in the fourth section. They suggest that international research networks: (1) rely on closely-knit communities and on a handful of coordinators; (2) exchange information or trust irrespective of the social and professional attributes of their members; (3) are negatively affected by distance and borders. The discussion section builds on our interviews to explain how research teams responded to the pandemic while the conclusion summarizes our main findings.

## 2. Literature review

### 2.1. Diversity and international team networks

The ability of teams to maintain and develop activities during major crises heavily depends on the internal composition or *diversity* of each network, defined as the "distribution of personal attributes among interdependent members of a work unit" (Jackson et al. 2003: 802). In addition to being embedded or playing the role of brokers, social actors are also connected to other nodes that may or may not share the same attributes, such as age, gender, or education. Actors may develop a homogeneous or diverse network, depending on whom they are connected, irrespective of their structural position. In Figure 1, for example, the diversity of actors to which actors B and C are connected varies greatly, despite them being in a similar structural position: while B is exclusively connected to peers who share the same attribute (black), C has developed a diverse network (black and white).

Figure 1. Social structure and diversity

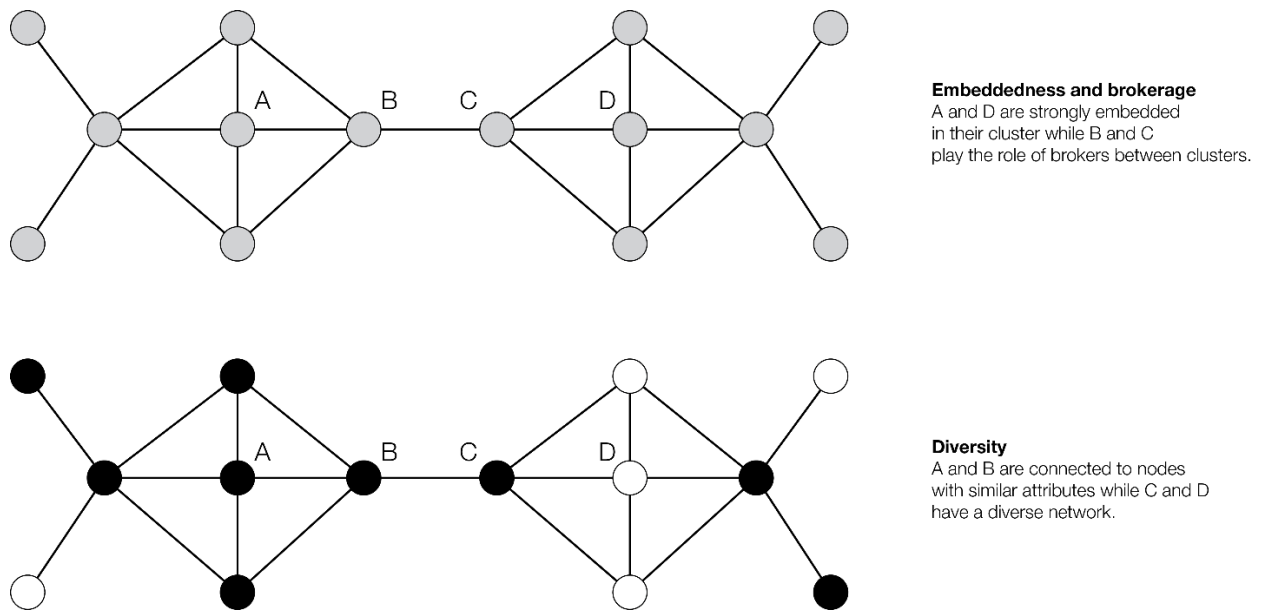

**Embeddedness and brokerage**
A and D are strongly embedded in their cluster while B and C play the role of brokers between clusters.

**Diversity**
A and B are connected to nodes with similar attributes while C and D have a diverse network.

Source: authors.

Extant literature has long debated the importance of diversity in social networks (Tilghman et al., 2021), highlighting two contrasting forces. On the one hand, network science emphasizes that social ties rarely form at random and, for this reason, tend to encourage homogeneity. People tend to be connected to peers who share similar attributes, a principle known as *homophily* (McPherson et al. 2001). On the other hand, collaborative networks operate in a highly uncertain environment in which diversity is key to anticipate, cope with, and adjust to changes (Duchek et al. 2020, Linkov et al., 2022). Watts and Koput (2014) show, for example, that a preference for diversity in tie formation increases the ability of networks to retain or restore their basic properties when key nodes or ties are removed, making diverse networks more resilient to external shocks. For this reason, diversity increases the likelihood of finding a job (Son and Lin, 2012), is positively associated with social and cultural capital (Tindall et al., 2012), increases performances when completing complex tasks (Maroulis et al., 2020), leads to innovation (Nielsen et al., 2017), and is positively correlated with a paper's impact and novelty (Yang et al., 2022).

## 2.2. Distance, borders, and international team networks

International research teams evolve in a challenging environment. Developing and maintaining relationships outside one's own institution is a long and costly process due to the propensity of social ties to decline sharply with distance (Onnela et al., 2011, Balland et al., 2022). In addition to being close to each other, scientists involved in collaborative projects must also overcome cultural, functional, and institutional differences that may affect how knowledge is produced and exchanged (Boschma, 2005; Lunquist and Trippl, 2013). Examining international collaborations between 193 countries, Vieira et al. (2022) show, for example, that researchers in countries that share a common language and colonial ties tend to publish more together, while a country's political regime or level of socioeconomic development have a marginal role.

A key challenge in this collaborative process is that new technologies have facilitated the exchange of codified information among researchers across various institutions, while knowledge production still relies on tacit information that is difficult to formalize and rooted in context (Pan et al., 2012). For this reason, physical distance remains one of the most constraining factors in knowledge production and exchange, even when researchers have access to the latest technology or share the same *lingua franca*. This is particularly true of international collaborations, in which the various obstacles associated with borders that can potentially limit exchanges between countries with different languages, institutions, and economic systems (Walther and Reitel, 2013; Cerina et al., 2014; Sohn et al., 2020). Verginer and Riccaboni (2020) show for example that even if scientists are among the most mobile professionals, their global mobility patterns are still marked by national borders. In Europe, international borders still shape collaborations between researchers (Benaim et al., 2016; Hjaltadóttir et al., 2020). Despite several decades of European research programs, the propensity to collaborate among scientists declines exponentially, while it declines as a power of distance in the fully integrated market of the United States (Cerina et al., 2014).

## 3. Data and methodology

Two related datasets are combined to model social networks: (1) an attribute table containing demographic information about who the nodes are, and (2) an adjacency matrix containing relational information about how the nodes are connected to each other. Because social actors embedded in a network are statistically dependent, network data are typically collected on whole populations or using nonprobability sampling techniques where actors identify other actors from among their acquaintances (Illenberger and Flötteröd, 2012).

*3.1. Data collection and structuring*

We surveyed three research networks connecting the University of Florida in the United States to Africa. These networks were selected for their interdisciplinary nature, their explicit international dimension, and their focus on public policies. Even more importantly, these three networks all successfully recovered from the impact caused by COVID-19, which represents a major achievement in the context of Africa (Vieira and Cerdeira, 2022), a continent where historically there has been less investment in scientific infrastructure and education.

- The Ghana Land Use Project (GALUP) was created in 2020 to help Ghanaian planners manage the linkages between deforestation, agriculture, and urbanization. Initially based on a series of face-to-face workshops organized in Ghana, the GALUP project was forced to design an online platform through which training could be conducted. These efforts paid off, as the team secured a major grant from NASA in 2022 to help monitor and assess artisanal mining and charcoal production in Ghana.
- The Sahel Research Group (SRG) was established in 2016 to promote social science research in the West African Sahel. Despite being heavily affected by travel restrictions, the team was able to publish an 800-page handbook on the Sahel in 2021 and secure a major grant from the Department of Defense in 2022 to study the perception of climate change in the Sahel.
- The Feed the Future Innovation Lab for Livestock Systems (LSIL) was created in 2015 to develop long-term research and capacity-development efforts in Africa. LSIL is composed of two major two structures: 1) a Management Entity which oversees research projects and 2) more than 30 unique research projects carried out in multiple countries in Sub-Saharan Africa and Asia. In 2020 when COVID-19 became a pandemic, the first round of projects carried out by LSIL were ending. Travel restrictions greatly limited the ability of members to travel to project sites and affected some of the funded projects' final activities. However, LSIL successfully applied for and received a second round of 5-year funding (2020-2025).

An online survey was conducted among the 22 researchers of the GALUP network and the 35 members of the SRG network to assess the level of information exchange and trust among them. 82% of the researchers of the GALUP network and 86% of the SRG network responded to our survey (Table 1). These high response rates allow us to minimize the number of missing nodes and ties that could affect the validity of the overall findings (Borgatti et al., 2006). All researchers in each research network had access themselves to relevant tools for international collaborations like video conferencing, virtual collaboration platforms, and project management software that are used to manage geographically-dispersed teams.

The first part of the survey collected demographic information on the name, race and ethnicity, country of origin, gender, highest degree obtained, discipline, and current employer of the actor, which was used as the attributes. By selecting three well established research networks that share a common language (English, and French when needed), we hope to minimize other factors such as language barriers between English and French-speaking countries or time zone differences between the US and Africa that may shape international networks. The second part of the survey asked the respondents to select the individuals they interact with the most through their research within each network. Direct interaction involved phone calls, video calls, face-to-face meetings, and emails sent between two persons.

Table 1. Response rate by survey

|  | Number of nodes | Number of nodes surveyed | Response rate |
|---|---|---|---|
| Ghana Land Use Project (GALUP) | 22 | 18 | 82% |
| Sahel Research Group (SRG) | 35 | 30 | 86% |
| Feed the Future Innovation Lab for Livestock Systems (LSIL) | 391 | 391 | 100% |

Source: authors

The respondents were also asked to select the individuals they would go to for advice on carrying out aspects of their research work. This advice could include information about the organization, research initiative, or scientific aspects of their work. For both the "information" and "trust" networks, a list of names including all the researchers employed on a permanent or temporary basis in the networks was provided to the respondents. The much larger LSIL network was surveyed using 79 co-authored peer-reviewed articles published from 2018-2021 by 391 researchers using annual reports. We also collected information about the discipline, education, employer, gender, and country of residence of each author using the personal and institutional website of each researcher.

Our next step was to create a matrix that contained the names of the researchers of each unit in rows and columns. In the GALUP and SRG networks, each of the researchers correspond to a node, and the links of the network represent information flows or trust between researchers. These networks are directed, since we don't presume symmetry or reciprocity between researchers. They are also unweighted: each cell is coded 1 if the researchers indicate that they interact with or trust other researchers, and 0 otherwise. In the LSIL network, each of the authors correspond to a node, and the links between them represent co-publications. The network is undirected because if one author has published with another, the reverse is also true, and weighted by the number of joint publications. The last step is to combine both the attribute table containing the properties of the actors and the matrices representing the ties between them (Figure 2).

Figure 2. International research networks by country

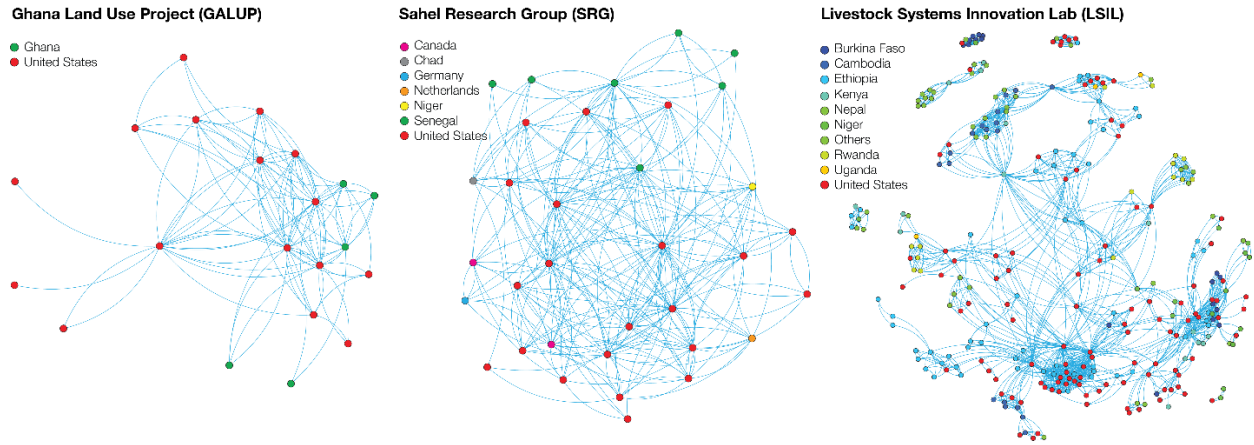

Source: authors. Each node represents a scientist and the ties represent information exchange between researchers (GALUP and SRG networks) or co-publications (LSIL).

The fact that the GALUP and SRG networks were collected using a survey, while the LSIL network was constructed from archival sources, can yield distinctively different network structures. Co-author networks will typically exhibit high clustering and transitivity because of how scientists communities are structured. In future work, Neal's (2022) R package could be used to focus on the backbone of the LSIL network and reduce the extent to which this network is distorted by the method used to collect the data.

## 3.2. Composition of international research networks

We assess the internal diversity of international research networks by using the concept of homophily, which assumes that social actors tend to be connected to peers who share similar attributes (McPherson et al., 2001). Our analysis determines whether researchers from the same disciplinary field, education level, university, place of origin and residence, and race or ethnicity are more likely to work together and trust each other. Homophily is assessed using the E-I index that measures the number of ties external to the groups minus the number of ties that are internal to the group divided by the total number of ties (Krackhardt and Stern, 1988). The E-I index coefficient ranges from +1 to -1. A large positive value of the E-I index indicates that connected nodes tend to possess very different properties (heterophily). A large negative value indicates that nodes associate with similar others (homophily). Values close to 0 mean no strong association of the property values between connected nodes.

We acknowledge that there is a natural imbalance in the network. For example, in the LSIL data, we observe that 37% of authors are from United States, 18% from Ethiopia and for 28 countries there are less than 1% of the authors. This means that people in the United States are more likely to work with people from the same country as well. We address this by permuting the network, and detect if most of the collaborations are only a result of proximity rather than other variables.

*3.4. Impact of distance and borders*

To measure the impact of geography on international research networks, we first run a regression analysis in which the dependent variable is the propensity to develop a social tie between two researchers. The independent variables are five attributes of the researchers: (1) the Euclidian distance between them in kilometers, (2) their highest degree achieved (high school, bachelor, masters, doctorate), (3) their employer, (4) their scientific discipline, and (5) their gender.

We consider two researchers (nodes) $i$ and $j$ and define the binary variable $W_{ij} = 1$ if they have worked together in the past and zero if they have not. For a network with $n$ nodes, there are $n(n-1)/2$ possible collaborations among the researchers. We construct a set of distance indicators between any pair of researchers, such that $D_{ij}^{di}$ is a binary variable, where $D_{ij}^{di} = 0$ if $i$ and $j$ belong to the same discipline and 1 if it is different. Similarly, we construct the distance in terms of gender, race, country of origin, country of residence, affiliation, and employer. Finally, we measure the Euclidean distance between $i$ and $j$, expressed as $D_{ij}$. This is crucial for collaborators who work across large countries (for example, people from the East and West Coast in the USA). Distances are exactly zero if $i$ and $j$ share the same attribute(s). A similar analysis is also considered for the trust between researchers, where we define the binary variable $T_{ij} = 1$ if they trust each other.

To detect how distance discourages working together or trusting others, we use an ordinary logistic regression. Although it should be expected that if A collaborates with B, and B collaborates with C, then A collaborates with C, we do not attempt to remove this effect but to quantify it and see how clusters of collaborations are formed. Logistic regressions are frequently used to analyze data where the outcome is a binary variable. In our case, we are interested in modelling whether two researchers work together (modelled as $W_{ij} = 1$) or not (modelled as $W_{ij} = 0$). The logistic regression accepts continuous variables (such as the distance between two researchers) but also binary variables (whether two researchers belong to the same institute or not) and other covariates. In our model, the probability of two researchers working together is expressed as:

$$P\big[W_{ij} = 1 \big| D_{ij}, D_{ij}^{di}, \ldots, D_{ij}^{gender}\big] = \frac{\exp\left(\beta_0 + \beta_1 D_{ij} + \beta_2 D_{ij}^{di} + \cdots + \beta_m D_{ij}^{gender}\right)}{1 + \exp\left(\beta_0 + \beta_1 D_{ij} + \beta_2 D_{ij}^{di} + \cdots + \beta_m D_{ij}^{gender}\right)}.$$

The coefficients $\beta_0, \beta_1, \ldots \beta_m$ are obtained through a regression between the collaborations in the network and the observed distances. If $\beta_k < 0$ for some distance variable, the probability that two researchers work together increases as they are closer to each other with respect to that dimension. Thus, the Null Hypothesis of interest is whether $H_0 = \{\widehat{\beta_k} = 0\}$.

To assess the importance of international borders on research networks, we build on Medina and Hepner's (2011) seminal work to test whether links between researchers are dependent on spatial clustering. We use the variable "is in the same country". We start by identifying clusters based on spatial distance, then we identify clusters based on social distance, and then compare the two. Based on the network that describes which researcher has collaborated with whom, we detect communities within that network using a measure of similarities between vertices based on random walks developed by Pons and Latany (2005). That means, we find densely connected subgraphs, or groups of researchers where most of them have worked together in the past. Researchers that have worked alone form their own community, but people with multiple collaborators are assigned to a group, or community. Each researcher is assigned to a unique community, even if there are overlaps between collaborators. Formally, forming communities means assigning researcher $i$ to community $C_j$, where most members of community $C_j$ are likely to have worked among each other and not with others. For example, for the LSIL Network with 391 researchers (nodes) and 1789 pairwise collaborations, there are 121 different communities detected, many of them with a single node (Figure 3).

Figure 3. Detecting communities within the LSIL network

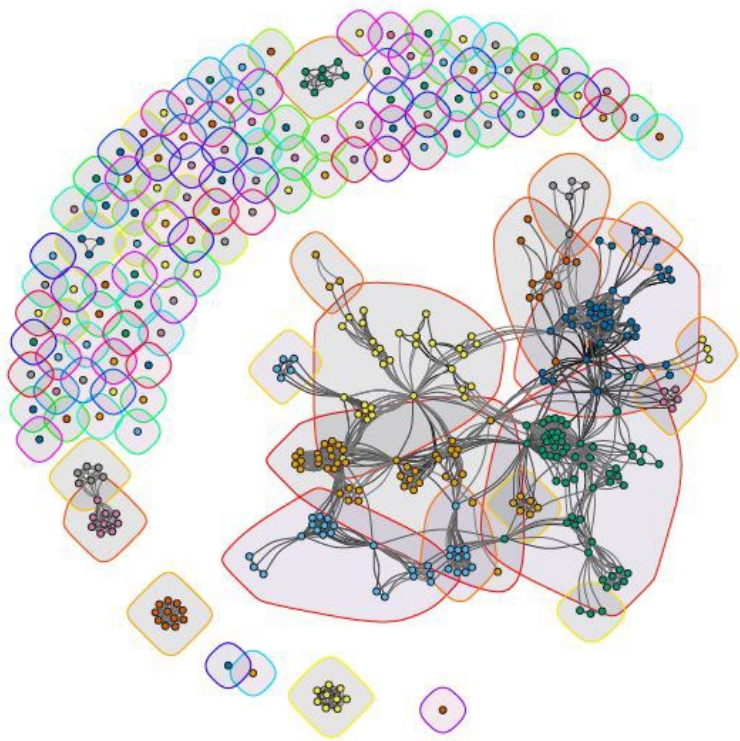

Source: authors. Note: each node represents a researcher. Ties represent co-publications. Communities are indicated with colors, We used short random walks to identify communities (Pons and Latapy, 2005). This technique separates individual researchers and communities that work closely with others. igraph (Csardi and Nepusz, 2006) in R was used to map communities.

These quantitative approaches are complemented with semi-structured interviews with 6 key researchers, three males and three females. For each team, we asked the Principal Investigator (PI) and one of the most active researchers to discuss the impact of the pandemic on scientific activities and the strategies developed by the team to respond to the pandemic. Qualitative assessments provide crucial information about the formation and temporal evolution of interpersonal relationships that cannot be accurately understood through social network analysis, that focus on the overall architecture of a network or the centrality of a node. The respondents were asked to discuss what were the biggest strengths of their team, how Covid-19 had affected the cohesion, interactions and functioning of their team, and how they adjusted when the pandemic first occurred.

## 4. Results

### 4.1. Social structure of international research networks

The structure of the three Africa-focused surveyed networks suggests that researchers rely on closely-knit communities, particularly the GALUP and the SRG networks, which exhibit a high density and numerous ties per actor (Table 2). Actors communicate with one another through a small number of steps and cluster together in small interdisciplinary clusters, as suggested by a low average path length (actors are separated by roughly two ties). This structure facilitates information exchange and trust between countries and across fields. As one of the core researchers of the Sahel Research Group explains: "Our biggest strength is the personal connections, the longstanding element of trust and respect and admiration of members towards one another. It is rather unique. And it is the size and reach of the network. So, any countries in the Sahel one goes, one can rely on those networks" (interview 3, 9/21/2022).

Table 2. Descriptive metrics pertaining to the three surveyed networks

|  | SRG | | GALUP | | LSIL |
|---|---|---|---|---|---|
| Type of network | Information | Trust | Information | Trust | Co-authorship |
| Nodes | 34 | 33 | 21 | 18 | 293 |
| Ties | 254 | 163 | 107 | 65 | 1445 |
| Density | 0.23 | 0.15 | 0.24 | 0.21 | 0.03 |
| Average degree | 14.94 | 9.88 | 10.00 | 7.22 | 11.51 |
| Clustering coefficient | 0.50 | 0.36 | 0.58 | 0.41 | 0.89 |
| Average path length | 1.94 | 2.14 | 1.84 | 1.83 | 3.99 |

Source: authors.

The structure of the LSIL co-authorship network is characterized by both high path length and clustering coefficient, typical of a regular network. The density is very low (3%) compared to the other networks (15-24%), which is due to the fact that it is larger. The average degree in this network is similar to the other networks. The network is rather cohesive: Only five subclusters are not connected to the main component. This is remarkable considering the large size of the research team. As the Director of LSIL notes "we have been able to build something remarkable. And a lot of that is that these people have not had egos. They've been willing to collaborate, willing to work together, travel together, face difficulties together" (interview 6, 10/6/2022). The network is structured around several subcommunities that tend to publish regularly with each other, which explains why the average number of steps needed to reach any actor of the network is quite high (3.99).

The analysis also suggests that the surveyed networks are polarized by a handful of actors who have established durable relationships both within and outside their country. The LSIL Director notes that the existence of such coordinators makes the network quite resilient: "When Covid happened… we had country coordinators at that time, they were country coordinators in some countries and regional coordinators in others. So these were all good scientists. And we had been very diligent in asking them to write out very robust proposals in terms of what they were planning to do and making sure that they were using the scientific method" (interview 6, 10/6/2022). Both the "information" and "trust" networks are organized in dense communities dominated by a handful of well-connected nodes. As a result, degree centralization, which measures the existence of actors with numerous ties, is particularly high, while betweenness centralization, which measures whether the network contains brokers, is low (Figure 4).

Figure 4. Centralization scores by research network

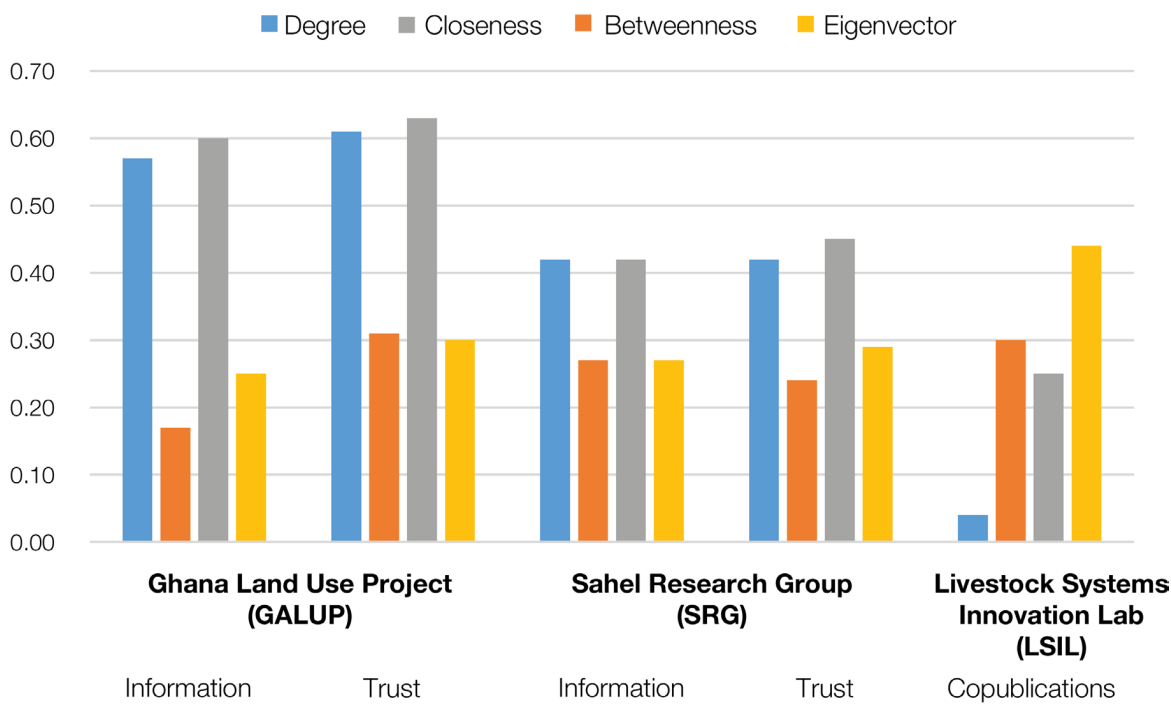

Source: authors. Note: the figure indicates to what extent each network is dominated by nodes that are connected to many others (degree), bridging different parts of the network (betweenness), close to the center of the network (closeness), or connected to well-connected nodes (eigenvector). Values close to 0 indicate that the network is rather decentralized while values close to 1 indicate that it is heavily centralized around several key nodes.

High eigenvector centralization scores in the LSIL network indicate that the most central authors publish with co-authors who also have numerous co-publications, an illustration of the principle of "preferential attachment" that argues that the most connected nodes are the most likely to receive new ties (Newman, 2001). The structural importance of coordinators is particularly visible when each node is represented according to the number of ties it has, as in Figure 5. In both the GALUP and SRG networks, the most central actors connected both to numerous colleagues in the U.S. and to foreign partners in Africa are the PI or Director of the network.

These links are strengthened by the interdisciplinary nature of both networks. The Coordinator of the GALUP project explains, for example, that "The purpose of this team is to find really interesting ways to combine remote sensing data and integrate that [with land use planning]. Remote sensing... is telling us from satellite, we can see how the land is changing over time. And we can tell the decision-makers: 'If you don't do anything, these trends are going to follow'" (interview 4, 9/27/2022). The SRG is also highly interdisciplinary. "We are from all sorts of backgrounds" explains its Director. "It ranges from the humanities to social sciences, to health and nutrition with a hard science component. So it is very interdisciplinary and I think that is also a big strength of the project" (interview 2, 9/15/2022).

Figure 5. Information exchange among the GALUP and SRG networks

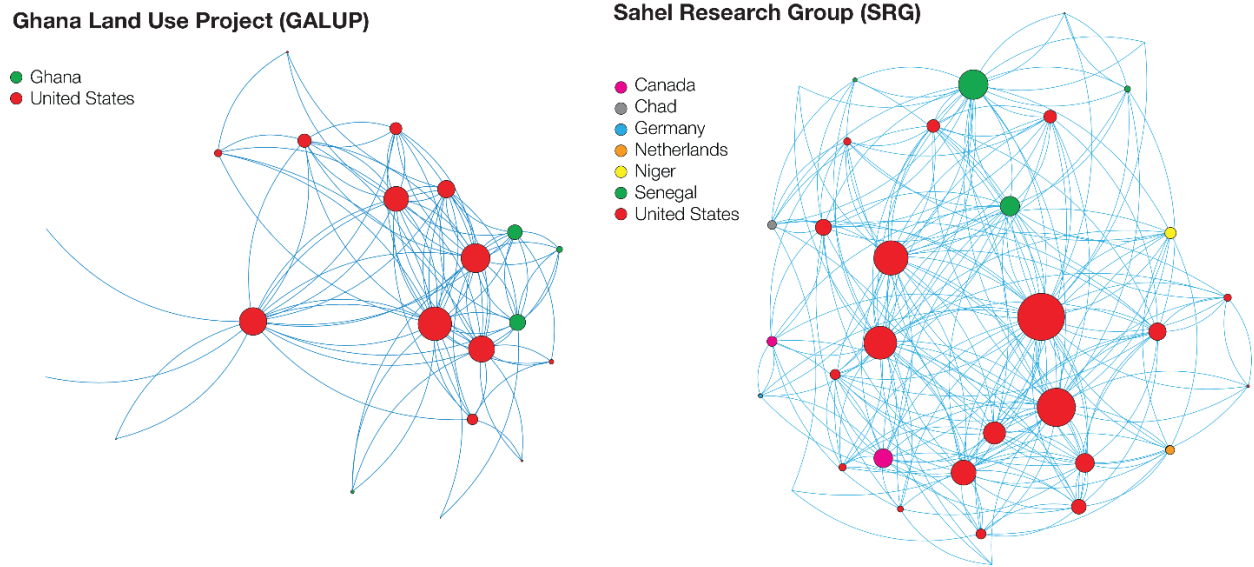

Source: authors. Note: each node represents a researcher and the ties between them indicate whether they tend to share information. The size of the nodes is proportional to the number of ties they have (degree centrality).

## 4.2. Composition of international research networks

Using the concept of homophily, we assessed whether researchers from the same disciplinary field, education level, place of origin and residence, and race or ethnicity were more likely to work together and trust each other. Our results presented in Figure 6 suggest that the surveyed networks tend to exchange information or trust each other irrespective of their social and professional attributes. The only exception is residence, for which homophily between the members of the GALUP and SRG networks can be observed: the E-I index that measures the proportion of ties external to the group minus the number of ties that are internal to the group is slightly negative. However, for most other variables, the largely positive index suggests heterophily, meaning that researchers are more likely to work with colleagues who belong to a different discipline, have a different level of education, place of origin, and race or ethnicity. Similar results were obtained for the LSIL network. diversity was emphasized by all the respondents. These results point to the benefits of building international research teams around a diverse pool of researchers. The LSIL Director notes, for example, that "We have a very, very diverse group, which is somewhat unique: diverse in terms of gender; diverse in terms of disciplinary background. I'm an animal scientist, [another colleague] is a vet, [another colleague] is an agricultural economist, [another colleague] is a sociologist. She does monitoring and evaluation. So we've had a very, very diverse team" (interview 6, 10/6/2022).

Figure 6. Homophily scores

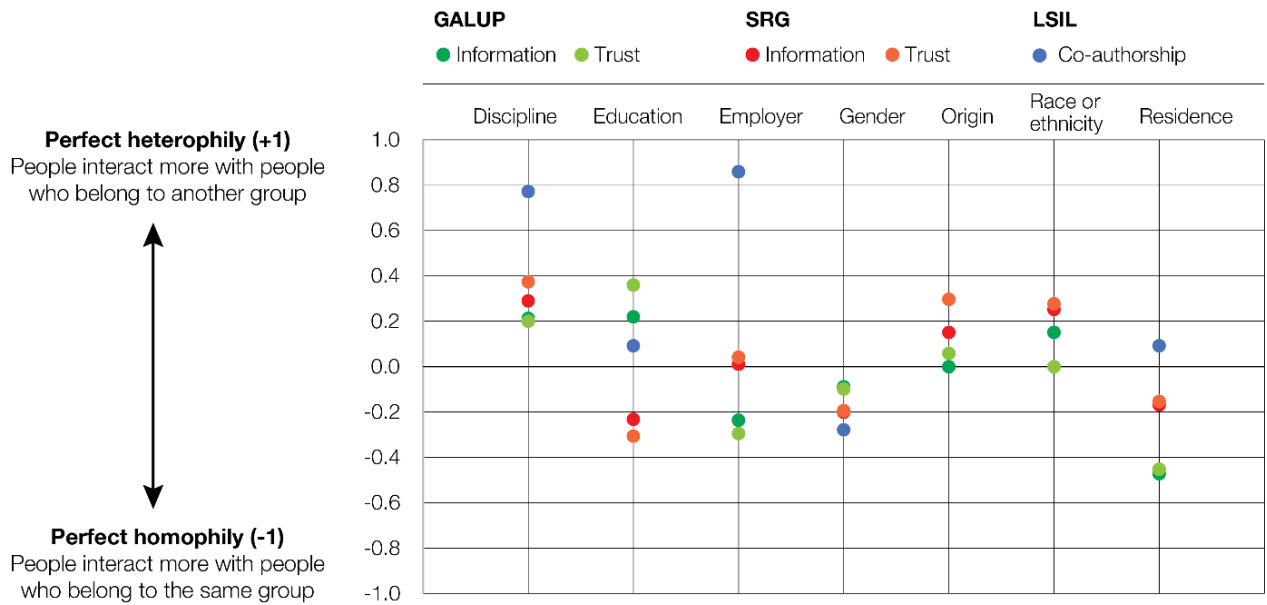

Source: authors. Note: the figure indicates whether researchers tend to exchange information, trust, or publish with researchers that share similar attributes. Date on place of origin and race/ethnicity were not collected for the LSIL network. Normalized scores are reported to consider the fact that the E-I index is restricted by the number of groups, relative group sizes, and total number of ties in a network.

*4.3. Impact of distance and borders*

The results of the regression model suggest that all factors reduce the probability that two people collaborate with others, with the exception of gender in the GALUP information network (Table 3).

Table 3. Regression coefficients

| | GALUP information | GALUP trust | SRG information | SRG trust | LSIL |
|---|---|---|---|---|---|
| Intercept | 0.414 | −0.103 | 0.953** | −0.372 | −1.145*** |
| | (−0.439) | (−0.526) | (−0.305) | (−0.360) | −0.058 |
| distance | −0.2916 | 0.0163 | −0.0597 | 0.0038 | −0.0065 |
| | (0.4033) | (0.6504) | (0.0252) | (0.0336) | (0.0050) |
| education | −0.507 | −1.075** | −0.427* | −0.587∗ | −0.443*** |
| | (−0.290) | (−0.349) | (−0.188) | (−0.239) | (−0.039) |
| employer | −0.851* | −1.468** | −1.136*** | −1.052*** | −1.562*** |
| | (−0.335) | (−0.534) | (−0.191) | (−0.251) | (−0.060) |
| gender | 0.126 | −0.076 | −0.171 | −0.048 | −0.310*** |
| | (−0.221) | (−0.287) | (−0.138) | (−0.181) | (−0.040) |
| discipline | −0.440 | −0.460 | −0.373 | −0.490 | −0.706*** |
| | (−0.367) | (−0.465) | (−0.265) | (−0.324) | (−0.048) |
| AIC | 520.011 | 353.247 | 1331.7 | 894.805 | 26482.489 |
| BIC | 552.333 | 385.569 | 1371.883 | 934.988 | 26551.997 |
| LogLikelihood | −252.005 | −168.623 | −657.850 | −439.403 | −13234.245 |
| Deviance | 504.011 | 337.247 | 1315.7 | 878.805 | 26468.489 |
| Number of obs. | 420 | 420 | 1122 | 1122 | 151710 |

Source: Authors. Note: ***p<0.001; **p<0.01; *p<0.05.

Table 3 suggests that physical distance and the existence of international borders reduce the probability that two people collaborate, trust, or publish together within the three surveyed networks. Two researchers are more likely to collaborate when they are from the same country if they are close to each other. Figure 7 shows, for example, that the probability of exchanging information or trusting each other decreases considerably after 1000 km for all networks, except for the SRG trust network. The LSIL network is the one for which the statistical relation between distance and work is the most significant. Considering two LSIL researchers from the same disciplines, gender, and degree, but with a different employer, we can estimate the probability that they collaborate depending on the distance between them. For distances of up to 5000 km, the probability is roughly 7%, meaning that one in every 14 pairs of researchers will work together. However, for a distance above 5,000 km, this probability decreases considerably.

Figure 7. Probability of cooperation and distance between researchers

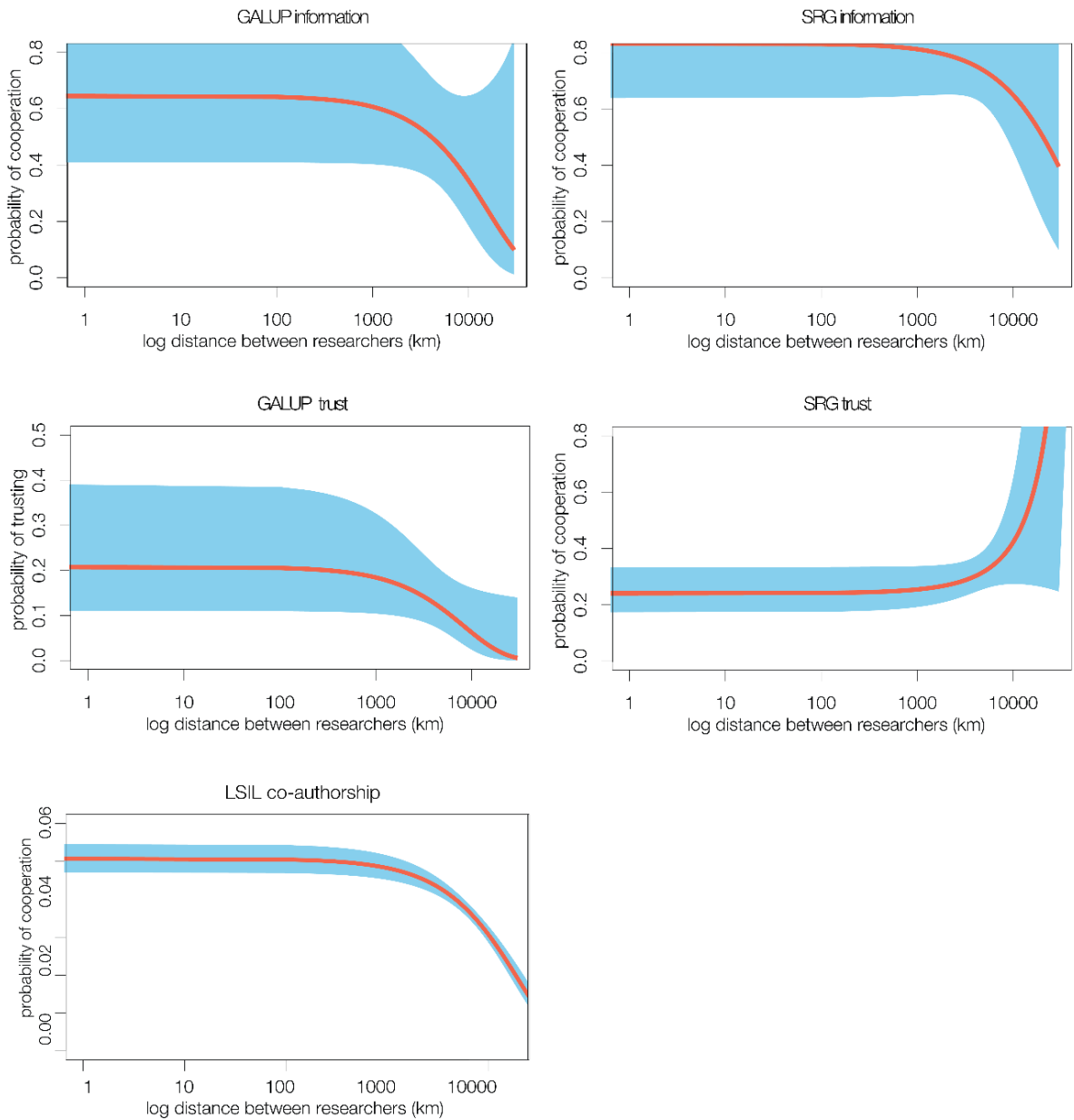

Source: authors. Note: the figure shows how the probability of exchanging information or trusting another researcher varies with Euclidian distance. The line corresponds to the modelled probability of exchanging information (vertical axis) given some distance between researchers (horizontal axis). The solid line is the estimated probability that two researchers collaborate, and the blue polygon corresponds to an interval constructed considering the standard error of the estimate. The LSIL network has the largest number of researchers, so the predicted interval is thinner than in the other two networks. For example, in the case of the SRG network, there are only a handful of researchers working in Africa and collaborating with people in North America, so distances between researchers above 10,000km only happens a limited number of cases.

We analyze the communities formed by the group of collaborators on each network. Given the different number of nodes and edges on each network, we obtain between 3 and 121 different communities. Each community is a group of researchers that have more collaborations between them than with others (Table 4). The LSIL network is composed of 121 communities including five communities with more than 20 researchers and 100 communities with only one researcher who published alone. The largest community of this network has 58 researchers, which represents nearly 15% of all the members of the Lab. The GALUP and SRG networks have fewer communities.

Table 4. Communities detected for the five collaboration networks

|  | GALUP | | SRG | | LSIL |
|  | work | trust | work | trust | |
| --- | --- | --- | --- | --- | --- |
| Communities | 7 | 10 | 3 | 8 | 121 |

Source: authors.

To establish whether international borders decrease the probability that two researchers collaborate, we measure the mean distance between the members of a community and compare this value with the one obtained for a random network with the same structure and number of communities. To ensure that our results are not affected by the result of a specific random permutation, we repeat the same process 1000 times, each time taking a different permutation (Figure 8).

Figure 8. Distances between individuals of the same community and permuted communities

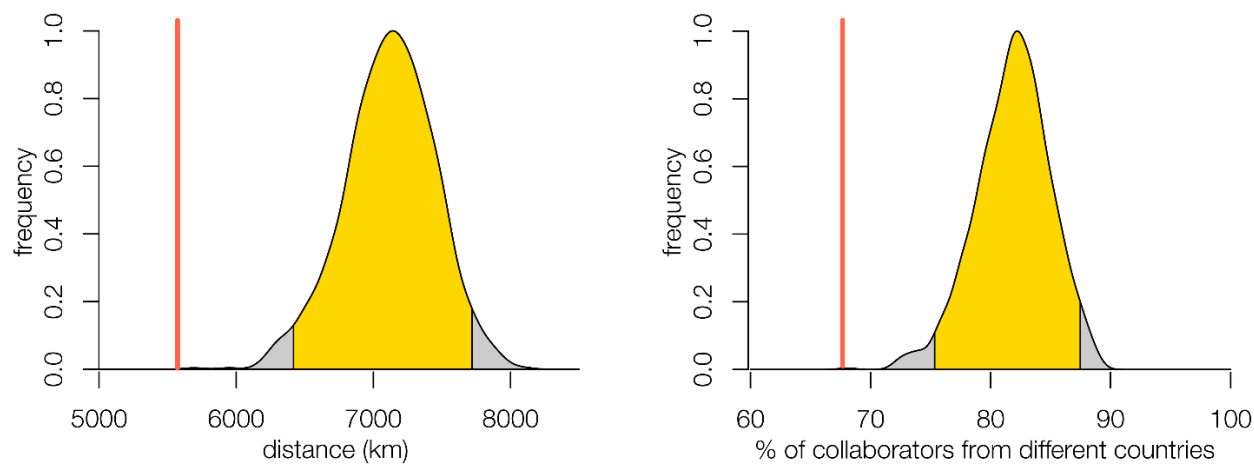

Source: authors. Note: The figure shows the distribution of nodes according to the distance between them (left) and to the share of researchers from different countries (right). The observed values are marked with a red vertical line. The 95% interval is colored yellow.

If the average distance within communities is smaller than the random value, it means that researchers are more likely to collaborate with colleagues located nearby. If the obtained value is larger than the random one, we can conclude that distance is not a limiting factor in scientific exchanges. Our analysis focuses on the LSIL network, which is considerably larger than the two other networks and allows a much more robust analysis of the role of borders on social ties. In the LSIL communities with more than one researcher, the average distance between researchers is 5572 km while the average distance between people is between 6397 and 7743 km when we assign individuals to a random community. In other words, collaborations within this network are happening with people that are 1530 km closer than what randomness would suggest, highlighting the key role of physical distance represents between them.

To measure the impact of international borders, a similar test is conducted for researchers living in a different country. In this case, we measure the frequency of pairs of individuals that belong to a community from a different country. In some communities, all researchers live in the same country, in which case the calculated frequency is 0%, and in other communities, all researchers live in different countries, in which case the observed value is 100%. On average, the results indicate that 68% of researchers of the same community are from the same country. We permute the communities as before and obtain that random communities are composed between 75 and 87% of researchers from different countries. Therefore, we conclude that communities are more likely formed inside the same country than internationally.

## 5. Discussion

The three international research networks selected in this study tend to adopt a structure in which researchers have a relatively high number of ties, are connected through short paths, and form dense communities. As one of the core researchers of the GALUP project notes, "The PI is a key person that really leads this project. She basically connected us to people in Ghana so that we have this communication back and forth" (interview 1, 9/14/2022). The ability to develop a balanced partnership with foreign countries is one of the key elements that explains the resilience of the surveyed teams in recent years. Activities in foreign countries are dependent on the trust relationships developed with local academic partners and stakeholders, especially when the joint activities include capacity building and co-publications, which both require numerous face-to-face interactions.

The LSIL Director illustrates this point by highlighting that his team has "incredibly strong buy-in among the government, the private sector, the civil society, and academia in the countries in our target countries. So, when COVID happened, and we could no longer travel, we were initially concerned. But then we were able to continue because… we had strong research partners in the country who knew what they were doing" (interview 6, 10/6/2022). This view is shared by another core member of the network, who notes that the strength of LSIL is to have targeted a small number of universities, livestock services and ministries with whom long-term relationships have been developed: "All our teams are working with target country institutions. So as much as the international partner was not able to travel, the teams in the field would still be able to do so" (interview 5, 9/13/2022).

None of the three teams surveyed in this paper could have succeeded if their views had been imposed, from above, on their African partners. Although the funding primarily comes from the United States, the ultimate objective of each team is to build long-term trust relationships that can survive the rather limited temporal scope of their joint projects in Africa. The GALUP Coordinator notes, for example that "there are local priorities which really need to be incorporated when you are doing scenario generation. And that's what we're doing. So that's what the goal of the project is: to really bring the stakeholders, remote sensing people, and land use planners together" (interview 4, 9/27/2022).

This long-term commitment to the region has led to a growing institutionalization of research initiatives, both in the United States where individual initiatives have transformed into well-established research groups, and in Africa, where punctual collaborations have served to launch more durable partnerships. As the SRG Coordinator recalls, "It started as an informal network, and then I think we have sort of formalized… From our University in the United States, we are building a much broader network" (interview 2, 9/15/2022). GALUP followed a similar evolution: it started as a project before becoming a research team within the University of Florida with its dedicated website, logo, and core researchers. While LSIL was never really a small project to start, the lab has nevertheless become more organized over time in how to do finances, monitoring, reporting, and other compliance related issues required by the management of large grants from USAID and the Bill & Melinda Gates Foundation.

So far, the growing institutionalization of the networks does not appear to have limited their flexibility to respond to ad hoc projects, as evidenced by the fact that all teams obtained substantial external funding during the COVID-19 crisis. For the SRG Coordinator, the flexibility of research networks is a comparative advantage over other more formal structures that contributes to explain its resilience: "Nobody is appointed in the Sahel Research Group, the group is a voluntary group… The group dynamic depends on people finding it very interesting to do so… I think that in many ways the fact that we had that collegial spirit and close collaboration, collegiality and friendship meant that we sort of rode out Covid" (interview 2, 9/15/2022).

The study also suggests that the surveyed networks rely on a diversified pool of researchers. None of the attributes that could contribute to fragmenting a network, such as gender, level of education, or origin, appear to be significant to explain how the network is structured. Diversity is widely recognized as one of the most important factors in the resilience of international activities by the surveyed teams: "The biggest strengths of LSIL are the teams of faculty" notes the LSIL Director. "This is a multidisciplinary, very diverse team, from the hard sciences to social sciences, food safety, gender, and livestock production disease" (interview 6, 10/6/2022). The SRG Coordinator takes a similar view on diversity, noting that of the eight faculty members involved in his group, only two were born in the U.S. "We're a pretty diverse group" he notes. "I'm always conscious about balancing that. Virtually everybody… who work in African Studies have diversity issues, racial equalities, and injustices built into their bones" (interview 2, 9/15/2022).

Echoing Uzzi's (1996) study on the consequences of embeddedness for the economic performance of organizations, respondents note that a delicate balance must be found between working with researchers who know each other well and introducing new research partners. The GALUP Coordinator was precisely in this situation when she responded to a call from SERVIR, a joint venture between NASA and USAID, that required a multidisciplinary and multinational team. "We have two women and four men on the team" she said. "Two are Europeans, two are Indians, one is Chinese, one is American. We are pretty diverse not only racially or gender wise, but we are also very diverse in our expertise. I cannot do what [my colleague] does. If [my colleague] was to drop out of the team, I don't think we have a project" (interview 4, 9/27/2022). Of particular importance in this project was to find researchers who had experience in the region or had worked in similar situations in a different context. In the absence of physical exchange between the PI in the United States and her partners in Ghana, these cultural brokers were able to reduce misunderstandings in the early stages of the project.

Distance and borders still represent a significant constraint on international networks. Our results suggest that the probability of exchanging information and trusting each other declines with distance. Researchers also tend to form communities segmented by country when it comes to publishing. These constraints, which have been documented in other studies (Cerina et al., 2014, Walther et al., 2021), are likely to remain significant among the global scientific community, despite the increasing use of communication technologies. They were amplified by the COVID-19 pandemic, which affected both the ability to meet in person within their country and to travel internationally.

All teams note that they initially felt helpless when confronted with the restrictions imposed by the pandemic. The LSIL Director, who coordinates research in 6 countries, notes that movement restriction, not COVID-19 itself, was the biggest challenge: "Our partners in Africa could not begin a project because they had to buy animals and the markets were closed down. They could not import products, because borders were closed" (interview 6, 10/6/2022). The GALUP Coordinator was in a similar position when she launched her new project in March 2020: "We were in our honeymoon period and the first trip that we were going to do [to Ghana] got cancelled. We had all these plans for capacity building, which were all in person. There was just nothing online at all" (interview 4, 9/27/2022). The disruption caused by COVID-19 was particularly brutal for those who were used to meeting regularly at work and socially. The SRG Coordinator explains that "For months I didn't have time to do anything anyway with the Sahel Research Group. I was in crisis management mode for a long time, like the whole university, for it feels like forever… It was a very big disruption" (interview 2, 9/15/2022).

Our interviews reveal that, after the initial shock caused by the pandemic, teams responded to the travel bans and border closure by emphasizing what they already did best. Known for its vibrant community, the SRG team compensated for the shrinking of geographic space by expanding its social network. If face-to-face meetings, fieldwork, and conferences were no longer possible, the idea was to invest further in social relationships that had already proved durable. The SRG Director explains that he drew even more on some of his network, inviting former graduate students to participate in the network activities: "We called them up and said: 'Could you join us for a Zoom meeting? Could you tell us about your country?' So that worked" (interview 2, 9/15/2022). SRG had the advantage of going into COVID-19 with a very coherent group of people working closely together. What was already a strong network of colleagues and friends became increasingly cohesive. "Relationships which survived Covid were those that were strong before" explains the coordinator. "If we are friends, we can deal with not seeing each other for six months and just do phone calls and see each other on Zoom... It works pretty well when you're working with the same person for ten years, but it's not the same when you have a brand-new colleague and you are meeting only in this circumstance" (interview 2, 9/15/2022).

Investing in preexisting relations was hardly an option for the GALUP team, which had been formed just before the pandemic and had never met its partners in Ghana. Instead, the team chose to replace physical interactions with new online tools, an alternative that may reflect the large number of engineers in the team. The GALUP Coordinator first asked herself what would be the most impactful decision her team could make given the situation. Her solution was to organize a series of workshops using an online training platform specifically designed for this project: "We had to find a way to move toward our goals… [Our stakeholders and us] were finding ways together and I think that explains why it worked… The first thing our stakeholders told us was 'We have never done this before'" (interview 4, 9/27/2022).

Being far larger than the two other teams, LSIL responded to the pandemic by further institutionalizing its relationships with its foreign partners and relying on the wide experience of its internal administration. For a network of nearly 400 people, investing in interpersonal relationships or developing new online tools was probably less relevant than consolidating its formal structure, both domestically and internationally. The Director describes this strategy by noting that one of the strengths of LSIL "is the team of staff we have had; we've had very strong cohorts, staff teams, who are very good at project management" (interview 6, 10/6/2022). His views are echoed by another core researcher of the network who notes that LSIL managed to survive the crisis and recover relatively rapidly due to its degree of institutionalization: "You can never say that the pandemic comes at the right time. Absolutely not. And I hope we don't have another one for a while. But we were established. It would have been extremely different if it had been in year 1 or 2 of LSIL" (interview 5, 9/13/2022).

## 6. Conclusion

This paper suggests that the overall structure and composition of international research teams is key to understanding how they respond to major external shocks such as the recent pandemic. The three networks surveyed in the paper tend to be organized around a small number of coordinators rather than dominated by a few brokers who connect the United States with foreign team leaders in African countries. This combination of arm's-length and embedded ties, which proves particularly critical for ad hoc teams that respond to calls for proposals launched by funding agencies. Research teams tend to be successful if they manage to create the right density of pre-existing ties: researchers who don't know each other prior to working together or who know each other too well are unlikely to create a research environment conducive to new ideas.

Diversity is therefore actively sought after by the surveyed teams, who note that a highly diverse pool of researchers can provide access to new ideas or resources that would not be available within their own discipline, university, or country. These results suggest a negative relationship between homophily and international collaborations. The paper also suggests that geographic distance and international borders remain a formidable obstacle to the establishment of fully integrated scientific networks, suggesting that researchers tend to form spatially close communities.

Despite obvious differences in size, objectives and leadership, the networks surveyed in the paper seem to have responded to the recent pandemic by adopting a similar strategy. Instead of developing radically new ideas, they invested even more in what made them strong in the pre-pandemic era, whether by focusing on their most trusted relationships, on their technical skills, or on their internal bureaucracy. While this strategy is quite understandable given the scale, speed, and disruptive consequences of the crisis, it also holds lessons for better understanding the evolution of research networks, beyond the three case studies considered in this paper. If the strategies adopted by successful international research teams depend on their initial strengths, perhaps the main implication of our results is that resilience should be considered as an evolutionary attribute, rather than a mere outcome, of a system.